# Coherent resonant transmission

H. S. Xu and L. Jin[*]
*School of Physics, Nankai University, Tianjin 300071, China*

The reflectionless coherent light transport in the coupled resonator array is investigated in the presence of intra-resonator intermodal coupling between the clockwise and counterclockwise modes, which plays a constructive role for modulating the light flow rather than inducing the unwanted backscattering. The interplay between the intra-resonator intermodal coupling and the inter-resonator couplings enables the coherent resonant transmission (CRT) of the properly superposed injection constituted by the clockwise and counterclockwise modes. The superposition coefficients of the initial excitation determine the mode chirality of the resonant transmission. Sequentially experiencing the time-reversal process of CRT and the CRT realizes the perfect mode conversion that the mode chirality of the injection wave switches into the opposite after resonant transmission. Our findings on the coherent light transport provide insights for the control and manipulation of light field in the integrated photonics, nanophotonics, chiral optics, and beyond.

*Introduction*. A coupled resonator optical waveguide (CROW) consists of a sequence of weakly coupled resonators [1]. The individual resonator can be a defect cavity in the photonic crystal [2], dielectric microdisk [3], microsphere [4], microcavity [5,6], or microring [7–13]. The tight-binding approximation well describes the CROW, where the light propagation depends on the evanescent coupling between adjacent resonators. The resonator in the CROW is an optical analog of the atom in the optical lattice; the modal amplitude of resonator plays the role of the wave function of atom. The CROW is a prominent platform for the simulation of many intriguing phenomena in quantum physics and condensed matter physics attributing to the feasible high-precision modulation and manipulation of light field including the construction of exceptional points [14,15], the photonic analogs of topological systems [16–21], and the edge mode lasing [22–26].

The light propagation in the CROW is usually disturbed by the defects and disorders caused by the fabrication imperfection. The undesirable surface roughness induces the unavoidable backscattering. In the CROW, the ring resonators support two degenerate modes: the clockwise (CW) mode and the counterclockwise (CCW) mode. The circling directions of photons define the opposite chiralities of the CW and CCW modes. The chirality plays the role of the pseudospin. The backscattering induces the modal mixing described by the intra-resonator intermodal coupling [27,28], which affects the resonance characteristics, causes the mode splitting, and limits the application performance [29]. The suppression of backscattering at low level is essential for many practical applications [30]; however, the mitigation of the defect-induced backscattering is extremely challenging. In contrast, the control of the backscattering-induced modal coupling via intentionally embedded scatterers is possible [31], which allows the delicate construction of the pseudo-spin-orbit interaction [32]. A pair of scatterers properly positioned in the resonator evanescent field enable the exceptional point and render many novel applications in non-Hermitian physics [33].

Coherent propagation is at the heart of many interesting phenomena including the coherent perfect absorption, reflection, and rotation in the wave transport and beyond [34–46], where the coherent superposition is extremely important. The coherent perfect absorber [34–39], as the time-reversed counterpart of a laser [47–49], is a celebrated example: The coherent injections superposed at proper phase and amplitude incoming from the opposite directions are perfectly absorbed as a consequence of the destructive self-interference. The coherent perfect absorption ubiquitously exists in lossy materials including the dissipative CROWs, and has wide applications in the wave tailoring, invisibility cloaking, and sensing [39].

In this work, we propose the coherent resonant transmission (CRT) in the CROWs: The coherent injection of the CW and CCW modes at the proper superposition of phases and amplitudes generates the reflectionless resonant transmission of the CW (CCW) mode. The CRT occurs at the interplay among the intra-resonator intermodal coupling, the inter-resonator coupling between the embedded resonators, and the inter-resonator coupling of the resonator array. Furthermore, the combination of the time-reversal process of CRT and the CRT realizes the perfect mode conversion. The chirality of the wave injection switches to its opposite after scattering. The proposed intriguing dynamics are implementable in many experimental platforms including the hybrid silicon microcavities and the dielectric microwave resonators.

---

[*]jinliang@nankai.edu.cn

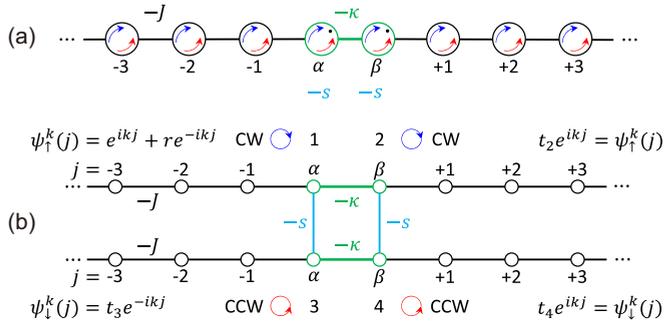

FIG. 1. Schematic of the CROW with two embedded resonators. (a) 1D coupled resonator array. The black dots in the central two resonators in green are the scatterers, which induce the intra-resonator intermodal couplings $-s$ in cyan. (b) Schematic of the quasi-1D tight-binding model of (a). The upper (lower) lead represents the CW (CCW) mode of the resonator array. The steady-state wave functions are marked.

*Modeling the CROW.* We consider that two coupled resonators $\alpha$ and $\beta$ are embedded in the middle of a resonator array as illustrated in Fig. 1(a). Both resonators have lossless scatterers or impurities as indicated by the black dots [50]. The resonators in the CROW are indirectly coupled through the linking resonators. The solid black and green lines represent the effective couplings between two neighbor resonators after adiabatically eliminating the light field of the linking resonators [17,18]. The resonators have an identical resonant frequency $\omega_c$ and support two degenerate modes, i.e., the CW mode (blue arrow) and the CCW mode (red arrow). The scatterers induce the energy exchange between the intra-resonator CW and CCW modes; thus, the CW and CCW light fields of the embedded resonators are coupled. In the coupled mode theory [28], the resonator array is modeled by a four-lead tight-binding model because of the scatterer-induced intermodal couplings of the embedded resonators [Fig. 1(b)]. The upper (lower) chain represents the CW (CCW) mode of the resonator array.

The resonator array $H = H_l + H_c + H_r$ consists of one scattering center and two leads. The lead Hamiltonians $H_l$ and $H_r$ are

$$H_l = -J \sum_{\sigma=\uparrow,\downarrow} \left( \sum_{j=-\infty}^{-1} a_{j-1,\sigma}^\dagger a_{j,\sigma} + a_{-1,\sigma}^\dagger a_{\alpha,\sigma} \right) + \text{H.c.}, \quad (1)$$

$$H_r = -J \sum_{\sigma=\uparrow,\downarrow} \left( \sum_{j=1}^{\infty} a_{j,\sigma}^\dagger a_{j+1,\sigma} + a_{1,\sigma}^\dagger a_{\beta,\sigma} \right) + \text{H.c.}, \quad (2)$$

where $a_{j,\sigma}^\dagger$ ($a_{j,\sigma}$) is the creation (annihilation) operator for the resonator $j$, the subscript $\sigma = \uparrow$ ($\downarrow$) indicates the CW (CCW) mode in the upper (lower) lead, and H.c. means the Hermitian conjugation counterparts. The scattering center includes four semi-infinite leads uniformly coupled as indicated by the solid black lines at the strength $-J$, which are distinguished by their (left/right) positions and their (CW/CCW) modes. The scattering center Hamiltonian $H_c$ reads

$$H_c = -\kappa \sum_{\sigma=\uparrow,\downarrow} a_{\alpha,\sigma}^\dagger a_{\beta,\sigma} - s \sum_{m=\alpha,\beta} a_{m,\uparrow}^\dagger a_{m,\downarrow} + \text{H.c.}, \quad (3)$$

where $-\kappa$ as indicated by the solid green line is the inter-resonator coupling between resonators $\alpha$ and $\beta$, and $-s$ as indicated by the solid cyan line is the intra-resonator intermodal coupling between the CW and CCW modes, which is delicately controlled by the scatterer [31,51].

*Light transport in the CROW.* The light transport is determined from the steady state of the effective four-lead structure. The steady state is a superposition of the plane waves $e^{\pm ikj}$ propagating in two opposite directions, where $k$ is the dimensionless wave vector and $j$ is the resonator site number.

We consider the CW mode excitation input from the left. This corresponds to the wave incidence in the upper left chain. The CW mode wave function for the upper lead is denoted as $\psi_\uparrow^k$ and the CCW mode wave function for the lower lead is denoted as $\psi_\downarrow^k$. The wave functions are set as $\psi_\uparrow^k(j) = e^{ikj} + re^{-ikj}$ for $j < 0$ and $\psi_\uparrow^k(j) = t_2 e^{ikj}$ for $j > 0$; $\psi_\downarrow^k(j) = t_3 e^{-ikj}$ for $j < 0$ and $\psi_\downarrow^k(j) = t_4 e^{ikj}$ for $j > 0$. Notably, $r$ is the reflection coefficient for the output in the upper left lead and represents the CW mode reflection; $t_2$ is the transmission coefficient for the output in the upper right lead and represents the CW mode transmission; $t_3$ is the transmission coefficient for the output in the lower left lead and represents the CCW mode reflection; and $t_4$ is the transmission coefficient for the output in the lower right lead and represents the CCW mode transmission.

The steady-state equations of motion for the sites on the uniformly coupled leads are given by $-J\psi_{\uparrow(\downarrow)}^k(j-1) - J\psi_{\uparrow(\downarrow)}^k(j+1) = E\psi_{\uparrow(\downarrow)}^k(j)$. Substituting the wave function in the above equation, the dispersion relation $E = -2J\cos k$ is obtained. The steady-state equations of motion for the scattering center $H_c$ are

$$-J\psi_\uparrow^k(-1) - \kappa \psi_\uparrow^k(\beta) - s\psi_\downarrow^k(\alpha) = E\psi_\uparrow^k(\alpha), \quad (4)$$

$$-\kappa \psi_\uparrow^k(\alpha) - J\psi_\uparrow^k(1) - s\psi_\downarrow^k(\beta) = E\psi_\uparrow^k(\beta), \quad (5)$$

$$-J\psi_\downarrow^k(-1) - \kappa \psi_\downarrow^k(\beta) - s\psi_\uparrow^k(\alpha) = E\psi_\downarrow^k(\alpha), \quad (6)$$

$$-\kappa \psi_\downarrow^k(\alpha) - J\psi_\downarrow^k(1) - s\psi_\uparrow^k(\beta) = E\psi_\downarrow^k(\beta). \quad (7)$$

The wave functions for the sites $|-1\rangle_\uparrow$, $|1\rangle_\uparrow$, $|-1\rangle_\downarrow$, and $|1\rangle_\downarrow$ are $\psi_\uparrow^k(-1) = e^{-ik} + re^{ik}$, $\psi_\uparrow^k(1) = t_2 e^{ik}$, $\psi_\downarrow^k(-1) = t_3 e^{ik}$, and $\psi_\downarrow^k(1) = t_4 e^{ik}$. From the wave function continuity, we obtain the wave functions of the scattering centers $|\alpha\rangle_\uparrow$, $|\beta\rangle_\uparrow$, $|\alpha\rangle_\downarrow$, and $|\beta\rangle_\downarrow$ in the form of $\psi_\uparrow^k(\alpha) = 1 + r$, $\psi_\uparrow^k(\beta) = t_2$, $\psi_\downarrow^k(\alpha) = t_3$, and $\psi_\downarrow^k(\beta) = t_4$. Substituting the wave functions into the steady-state equations of motion, the scattering coefficients are obtained [52],

$$r = \frac{(J^2 - \Omega^2)(J^2 e^{-ik} - \Omega^2 e^{ik}) - 4J^2 s^2 \cos k}{[4J^2 s^2 - (J^2 e^{-ik} - \Omega^2 e^{ik})^2]e^{-ik}}, \quad (8)$$

$$t_2 = \frac{2i\kappa J \sin k (J^2 e^{-ik} - \Omega^2 e^{ik})}{[4J^2 s^2 - (J^2 e^{-ik} - \Omega^2 e^{ik})^2]e^{-ik}}, \quad (9)$$

$$t_3 = \frac{2isJ \sin k (J^2 e^{-ik} + \Omega^2 e^{ik})}{[4J^2 s^2 - (J^2 e^{-ik} - \Omega^2 e^{ik})^2]e^{-ik}}, \quad (10)$$

$$t_4 = \frac{4is\kappa J^2 \sin k}{[4J^2 s^2 - (J^2 e^{-ik} - \Omega^2 e^{ik})^2]e^{-ik}}, \quad (11)$$

where we set $\Omega^2 = \kappa^2 - s^2$ in the expressions.

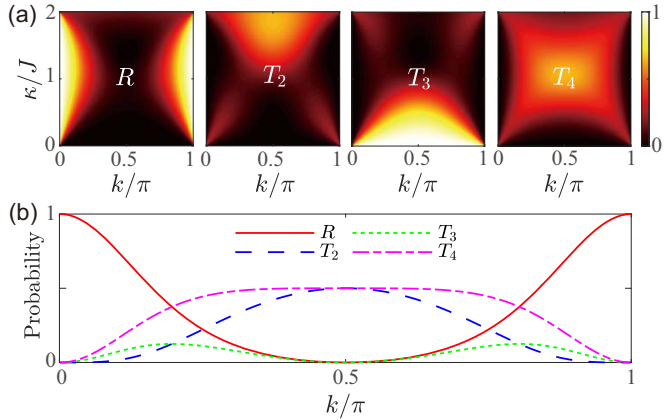

FIG. 2. Reflection and transmission $R = |r|^2$, $T_2 = |t_2|^2$, $T_3 = |t_3|^2$, and $T_4 = |t_4|^2$ for the situation $s = J$. (a) Contour plots. (b) $\kappa = \sqrt{2}J$.

Notably, the reflectionless condition $r = t_3 = 0$ requires the resonant condition $J^2 = \Omega^2$ and $k = \pi/2$. To be precise, when the inter-resonator coupling $\kappa$ and the intermodal modal coupling $s$ satisfy

$$J^2 = \kappa^2 - s^2, \quad (12)$$

the scattering coefficients for the input with the momentum $k = \pi/2$ are

$$r = t_3 = 0, \quad t_2 = iJ/\kappa, \quad t_4 = -s/\kappa. \quad (13)$$

This implies that the CW mode resonant input from the left penetrates the scattering center toward the right without reflection, and the transmitted waves consist of the CW and CCW modes with an amplitude ratio $J/s$ and a relative phase $e^{-i\pi/2}$.

Without loss of generality, the analytical reflection and transmission probabilities as functions of the coupling $\kappa$ and the momentum $k$ for a concrete case $s = J$ are plotted in Fig. 2(a). The reflectionless propagation occurs when the coupling strength is $\kappa = \sqrt{2}J$ at the resonant input $k = \pi/2$ as indicated in Fig. 2(b) and predicted in Eq. (12).

*Symmetry protected light transport.* The equivalent four-lead scattering system is mirror-reflection symmetric in both the horizontal and vertical directions. The reflection and transmission coefficients of waves incident from different leads are straightforwardly obtained from the symmetry protection. We denote the incoming wave as $\Psi_{\text{in}}$ and the outgoing wave as $\Psi_{\text{out}}$. The input and output are connected by the scattering matrix $S$ [53],

$$\Psi_{\text{out}} = S\Psi_{\text{in}}. \quad (14)$$

The scattering matrix $S$ completely characterizes the properties of the resonator array.

The CW mode wave incoming from the left is the input in the lead 1; the elements in the scattering are $S_{11} = r$, $S_{21} = t_2$, $S_{31} = t_3$, and $S_{41} = t_4$. The CW mode wave incoming from the right is the input in the lead 2. From the vertical mirror-reflection symmetry, the outputs are the scattering results of the input in the lead 1 if we exchange the outputs in the leads 1 and 2 and exchange the outputs in the leads 3 and 4; thus, we have $S_{12} = S_{21} = t_2$, $S_{22} = S_{11} = r$, $S_{32} = S_{41} = t_4$, and $S_{42} = S_{31} = t_3$.

The CCW mode wave incoming from the left is the input in the lead 3. From the horizontal mirror-reflection symmetry, the outputs are the scattering results of the input in the lead 1 if we exchange the outputs in the leads 1 and 3 and exchange the outputs in the leads 2 and 4; thus, we have $S_{13} = S_{31} = t_3$, $S_{23} = S_{41} = t_4$, $S_{33} = S_{11} = r$, and $S_{43} = S_{21} = t_2$.

The CCW mode wave incoming from the right is the input in the lead 4. From the horizontal mirror-reflection symmetry, the outputs are the scattering results of the input in the lead 2 if we exchange the outputs in the leads 1 and 3 and exchange the outputs in the leads 2 and 4; thus, we have $S_{14} = S_{32} = t_4$, $S_{24} = S_{42} = t_3$, $S_{34} = S_{12} = t_2$, and $S_{44} = S_{22} = r$.

From the mirror-reflection symmetry protection as listed above, the scattering matrix of the CROW reads

$$S = \begin{pmatrix} r & t_2 & t_3 & t_4 \\ t_2 & r & t_4 & t_3 \\ t_3 & t_4 & r & t_2 \\ t_4 & t_3 & t_2 & r \end{pmatrix}. \quad (15)$$

*Coherent resonant transmission.* The dynamics in the resonator array are protected by the time-reversal symmetry. The time-reversal process of a single CW or CCW mode input split into two transmitted waves of the CW and CCW modes is the CRT: two coherent input waves of the CW and CCW modes forming a reflectionless single CW or CCW mode output. The relative phases and amplitudes of the two transmitted waves determine the superposition coefficients for the CRT. To generate a resonantly transmitted single CW (CCW) mode output, the relative amplitudes of the CW and CCW modes in the input should be $J : s$ ($s : J$).

We demonstrate the CRT in the resonator array at the reflectionless condition $J^2 = \kappa^2 - s^2$ for the resonant input $k = \pi/2$. From Eq. (13), the scattering matrix is

$$S = (iJ/\kappa)\sigma_0 \otimes \sigma_x - (s/\kappa)\sigma_x \otimes \sigma_x. \quad (16)$$

The scattering matrix satisfies $S^{-1} = S^*$ from the unitary feature $S^{-1} = S^\dagger$ and the reciprocity $S = S^T$ [53]. We show the analytical results and perform the numerical simulations as the verification.

The initial excitation is a normalized Gaussian wave packet with the wave vector $k_c$ centered at the site $N_c$,

$$|\phi(0)\rangle = \Lambda_0^{-1/2} \sum_j e^{-(j-N_c)^2/(2\sigma^2)} e^{ik_c j} |j\rangle. \quad (17)$$

The center $N_c$ has the maximal intensity, $2\sqrt{\ln 2}\sigma$ characterizes the full width at half maximum, $v = 2J \sin k_c$ is the propagation velocity, and $\Lambda_0$ normalizes $||\phi(0)\rangle|^2 = 1$. The time evolution of the Gaussian wave packet in the resonator array is $|\phi(t)\rangle = e^{-iHt}|\phi(0)\rangle$.

From $\Psi_{\text{out}} = S(0, 0, 0, 1)^T = (-s/\kappa, 0, iJ/\kappa, 0)^T$, the single CCW mode initial excitation input from the right forms the CW and CCW modes transmitted wave outgoing toward the left at the ratio $s : -iJ$. Applying the time-reversal operation, the superposition of the CCW and CW modes at the ratio $s : iJ$ input from the left generates a reflectionless resonantly transmitted CW mode wave outgoing toward the right as verified from the simulation in Fig. 3(a) and predicted by

$$\Psi_{\text{out}} = S(-iJ/\kappa, 0, -s/\kappa, 0)^T = (0, 1, 0, 0)^T. \quad (18)$$

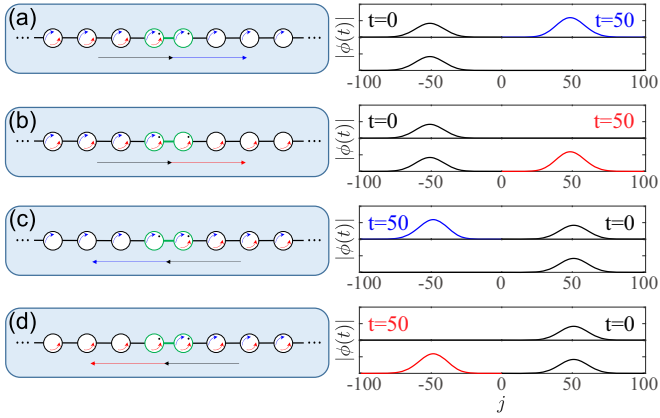

FIG. 3. Schematic and numerical simulation of CRT for $s = J$, $\kappa = \sqrt{2}J$. The CW mode and CCW mode in the initial excitation are superposed at the ratio (a) $i : 1$, (b) $1 : i$, (c) $i : 1$, (d) $1 : i$. The black arrows and black wave packets represent the coherent input; the blue (red) arrows and blue (red) wave packets represent the CW (CCW) mode resonant transmission. The initial Gaussian wave packets at $t = 0$ have $\sigma = 10$; $k_c = \pi/2$ in (a), (b) and $k_c = -\pi/2$ in (c), (d).

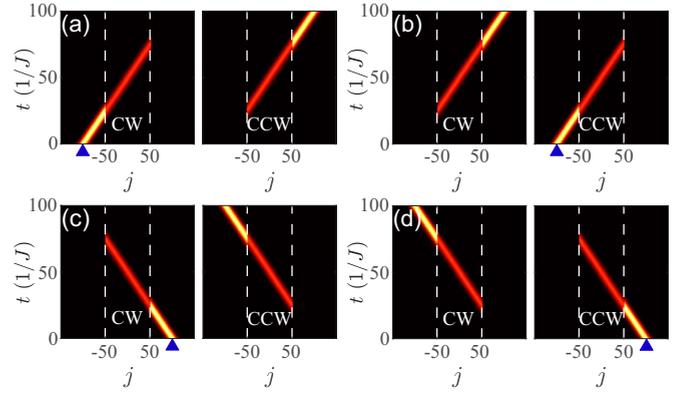

FIG. 4. Numerical simulations of perfect mode conversion for $s_1 = s_2 = J$, $\kappa_1 = \kappa_2 = \sqrt{2}J$. The intensities of the CW and CCW modes are separately depicted on the left and right panels of each plot. The time $t$ is in the unit of $1/J$. (a) Left input CW mode, (b) left input CCW mode, (c) right input CW mode, and (d) right input CCW mode. The initial Gaussian wave packets at $t = 0$ have $\sigma = 10$ centered at $N_c = -100$ with $k_c = \pi/2$ in (a), (b); and centered at $N_c = 100$ with $k_c = -\pi/2$ in (c), (d). The dashed white lines indicate the positions of the coupled defective resonators.

From $\Psi_{\rm out} = S(0, 1, 0, 0)^T = (iJ/\kappa, 0, -s/\kappa, 0)^T$, the single CW mode initial excitation input from the right forms the CW and CCW modes transmitted wave outgoing toward the left at the ratio $-iJ : s$. Applying the time-reversal operation, the superposition of the CCW and CW modes at the ratio $iJ : s$ input from the left generates a reflectionless resonantly transmitted CCW mode wave outgoing toward the right as verified from the simulation in Fig. 3(b) and predicted by

$$\Psi_{\rm out} = S(-s/\kappa, 0, -iJ/\kappa, 0)^T = (0, 0, 0, 1)^T. \quad (19)$$

Applying the vertical mirror-reflection operation to Eq. (18) yields

$$\Psi_{\rm out} = S(0, -iJ/\kappa, 0, -s/\kappa)^T = (1, 0, 0, 0)^T. \quad (20)$$

The superposition of the CW and CCW modes at the ratio $iJ : s$ input from the right results in the CW mode transmitted wave outgoing toward the left as verified from the simulation in Fig. 3(c). We also have

$$\Psi_{\rm out} = S(0, -s/\kappa, 0, -iJ/\kappa)^T = (0, 0, 1, 0)^T. \quad (21)$$

The superposition of the CW and CCW modes at the ratio $s : iJ$ input from the right results in the CCW mode transmitted wave outgoing toward the left as verified from the simulation in Fig. 3(d).

*Perfect mode conversion.* In a single resonator, the mode conversion was generated through unidirectional intermodal coupling between the CW and CCW modes [54–58]. In the CROW, the interference between the CW and CCW modes creates the perfect mode conversion.

The perfect mode conversion is possible for sequently embedded two setups of coupled defective resonators that support the CRT at $J^2 = \kappa_1^2 - s_1^2 = \kappa_2^2 - s_2^2$. The output $\Psi_{\rm out}$ in the resonator array after scattering is

$$\Psi_{\rm out} = S_2(\sigma_0 \otimes \sigma_x)S_1 \Psi_{\rm in}. \quad (22)$$

The term $\sigma_0 \otimes \sigma_x$ is inserted between $S_1$ and $S_2$ because the middle section between the two setups of coupled defective resonators stands for different leads in the two scattering processes. The middle section stands for the leads 2 and 4 in the first (second) scattering process, but stands for the leads 1 and 3 in the second (first) scattering process for the left (right) input.

From Eq. (16), the scattering matrix in Eq. (22) is obtained as

$$S_2(\sigma_0 \otimes \sigma_x)S_1 = \frac{-iJ(s_1 + s_2)}{\kappa_1 \kappa_2} \sigma_x \otimes \sigma_x + \frac{s_1 s_2 - J^2}{\kappa_1 \kappa_2} \sigma_0 \otimes \sigma_x. \quad (23)$$

Under the additional reflectionless condition

$$s_1 s_2 = J^2, \quad (24)$$

the scattering matrix reduces to $S_2(\sigma_0 \otimes \sigma_x)S_1 = -i\sigma_x \otimes \sigma_x$, which indicates the perfect mode conversion.

The chirality of the single mode injection wave switches between the CW and CCW modes after resonantly transmitted the scattering centers. The output $\Psi_{\rm out} = -i(0, 0, 0, 1)^T$ only exists in the lead 4 for the input $\Psi_{\rm in} = (1, 0, 0, 0)^T$ in the lead 1 and vice versa; the output $\Psi_{\rm out} = -i(0, 0, 1, 0)^T$ only exists in the lead 3 for the input $\Psi_{\rm in} = (0, 1, 0, 0)^T$ in the lead 2 and vice versa. The resonantly transmitted single mode output wave has the opposite mode chirality to that of the single mode input wave. The proposed system perfectly exchanges the CW and CCW modes for an arbitrary wave incidence.

We perform the numerical simulations of the perfect mode conversion in Fig. 4. The Hamiltonian of the resonator array has two copies of the coupled resonators $\alpha$ and $\beta$ respectively embedded at the positions $-50$ and $50$ of the resonator array. The total size of the resonator array is 300. The equivalent tight-binding system is a four-lead scattering system similar to that illustrated in Fig. 1(b); however, two rectangular structures of the sites $\alpha$ and $\beta$ present and space 100 sites.

The center of the initial Gaussian packet $N_c$ is located at the site $|-100\rangle_\uparrow$ ($|-100\rangle_\downarrow$) and $|100\rangle_\uparrow$ ($|100\rangle_\downarrow$) as indicated by the solid triangular markers in blue to simulate the initial CW (CCW) mode injection on the left and right sides, respectively. The wave vector for the Gaussian wave packet is $k_c = \pi/2$ for the wave propagation from the left to the right; and the wave vector for the Gaussian wave packet is $k_c = -\pi/2$ for the wave propagation from the right to the left. The input has the resonant frequency $\omega_c$ and the velocity is $2J$.

In the time interval from $t = 0$ to $t = 50$, the incident wave packet is scattered by the embedded scattering center at the site $j = -50$ ($j = 50$), and the wave packet splits into two transmitted wave packets of the CW and CCW modes propagating in the same direction. These dynamics confirm the time-reversal process of the CRT. In the time interval from $t = 50$ to $t = 100$, the two wave packets pass through the same scattering center at the site $j = 50$ ($j = -50$) and merge into a single transmitted wave packet without reflection. These confirm the dynamics of the CRT. The chirality of the initial injection wave at time $t = 0$ is opposite to the chirality of the final transmitted wave after passing through the two scattering centers at time $t = 100$; the mode chirality switch of the injected waves before and after the resonant transmission verifies the perfect mode conversion.

*Discussion and conclusion.* The coherent wave injection associated with the interference effect enables the intriguing applications including the coherent perfect absorption, reflection, and rotation that are absent for the incoherent wave injection [34,41,42,44–46]. We report the CRT and the perfect mode conversion as useful light transport phenomena induced by the intermodal coupling in the CROW.

The suppression of backscattering to enhance the resonator performance through weakening the resonator defect induced intermodal coupling is possible [59,60]. Alternatively, the CRT is proposed at the interplay between the intermodal coupling and the resonator couplings under the resonant condition. The intermodal coupling plays a pivotal role for the control and manipulation of light field instead of creating undesirable backscattering. Previously, only the robust edge mode transport in the topological CROWs is backscattering-immune [9,61].

The CRT ubiquitously exists in the CROWs with defective resonators under the resonant condition for the coherent wave input of the properly superposed CW and CCW modes and creates the reflectionless chiral light output of the CW or CCW mode [62,63]. The mode chirality of resonant transmitted output depends on the superposition coefficients of the input. Combining the time-reversal process of CRT and the CRT realizes the perfect mode conversion, where the chirality of injection wave switches between the CW and CCW modes. The intriguing dynamics proposed through constructively utilizing the intermodal coupling are insightful and would benefit the light flow molding in the integrated photonics, nanophotonics, chiral optics, and beyond.

*Acknowledgment.* This work was supported by the National Natural Science Foundation of China (Grant No. 11975128).